\documentclass[submission, Proceedings]{SciPost}

\usepackage{amsmath}
\usepackage{amssymb}
\usepackage{xspace}

\newcommand{\cp}{\ensuremath{\mathcal{CP}}\xspace}
\newcommand{\SM}{{\text{SM}}}

\newcommand{\kgamma}{\ensuremath{\kappa_\gamma}\xspace}
\newcommand{\kg}{\ensuremath{\kappa_g}\xspace}

\newcommand{\cv}{\ensuremath{c_V}\xspace}
\newcommand{\ct}{\ensuremath{c_t}\xspace}
\newcommand{\cttilde}{\ensuremath{\tilde c_{t}}\xspace}

\hypersetup{
    colorlinks,
    linkcolor={red!50!black},
    citecolor={blue!50!black},
    urlcolor={blue!80!black}
}

\usepackage[bitstream-charter]{mathdesign}
\DeclareSymbolFont{usualmathcal}{OMS}{cmsy}{m}{n}
\DeclareSymbolFontAlphabet{\mathcal}{usualmathcal}

\begin{document}

\begin{center}{\Large \textbf{
Measuring single-top-associated Higgs production at the HL-LHC\\
}}\end{center}

\begin{center}
Henning Bahl
\end{center}

\begin{center}
DESY, Notkestraße 85,\\
D-22607 Hamburg, Germany \\
henning.bahl@desy.de
\end{center}

\begin{center}
\today
\end{center}

\definecolor{palegray}{gray}{0.95}
\begin{center}
\colorbox{palegray}{
  \begin{tabular}{rr}
  \begin{minipage}{0.1\textwidth}
    \includegraphics[width=22mm]{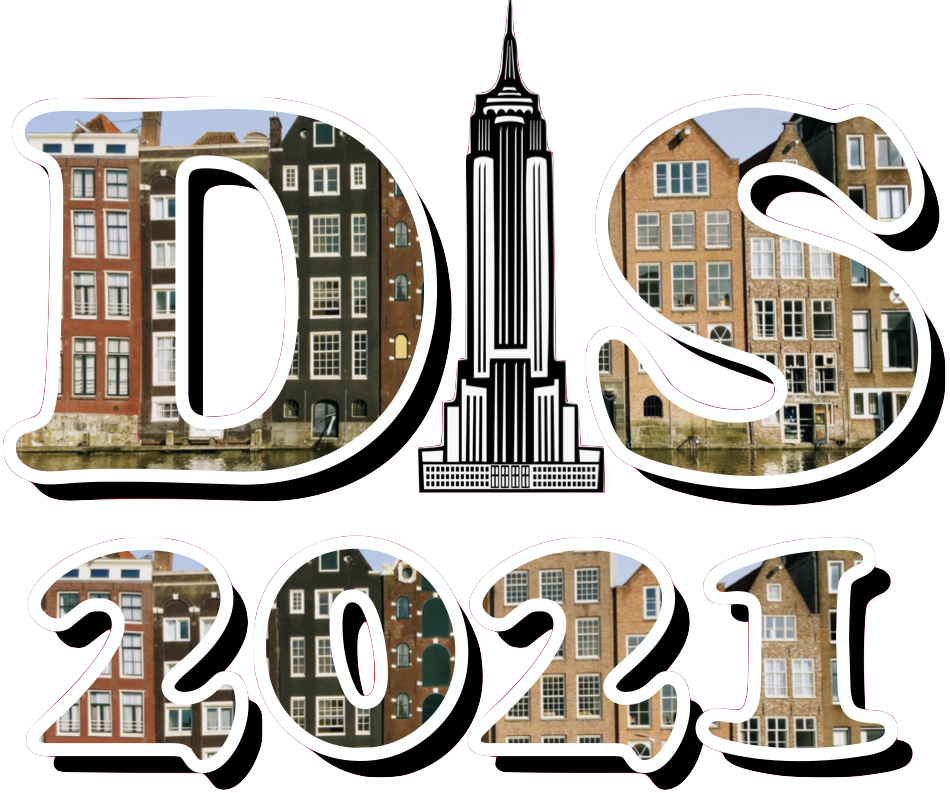}
  \end{minipage}
  &
  \begin{minipage}{0.75\textwidth}
    \begin{center}
    {\it Proceedings for the XXVIII International Workshop\\ on Deep-Inelastic Scattering and
Related Subjects,}\\
    {\it Stony Brook University, New York, USA, 12-16 April 2021} \\
    \doi{10.21468/SciPostPhysProc.?}\\
    \end{center}
  \end{minipage}
\end{tabular}
}
\end{center}

\section*{Abstract}
{\bf
Precision measurements of top-associated Higgs production are an important ingredient to unravel the \cp nature of the Higgs boson. In this work, we constraint the \cp nature of the top-Yukawa coupling taking into account all relevant inclusive and differential Higgs boson measurements. Based upon this fit, we show that it is crucial to disentangle single- and di-top-associated Higgs production for tightening indirect constraints on a \cp-odd top-Yukawa coupling in the future. In this context, we propose an analysis strategy for measuring $tH$ production at the HL-LHC without relying on assumptions about the Higgs \cp character.
}


\section{Introduction}

After the discovery of a new particle which is consistent with the predictions for the Standard Model (SM) Higgs boson, the investigation of this particle is one of the main tasks for future LHC and High-Luminosity LHC (HL-LHC) runs. Especially in light of the absence of any direct evidence for beyond the SM (BSM) physics, the precise determination of the Higgs boson's properties can give crucial hints to unravel the nature of BSM physics.

The determination of the Higgs boson's \cp properties is an important part of this program with a close connection to cosmology, since the Higgs boson could provide a new source of \cp violation allowing to explain the baryon asymmetry of the Universe (BAU). While experimental studies have already excluded the possibility of the Higgs boson being a pure \cp-odd state, the possibility of the Higgs boson being a \cp-admixed state is so far less constrained.

The focus of the present study~\cite{Bahl:2020wee} is the by magnitude largest Higgs--fermion--fermion interaction: the top-Yukawa interaction. While the \cp properties of the top-Yukawa coupling can also be constrained by electric dipole measurements or by demanding a sufficient amount of \cp violation to explain the BAU (see e.g.~\cite{Fuchs:2020uoc}), we concentrate on the constraints imposed by current and future LHC measurements.

At colliders, \cp-violating couplings can be constrained directly by measuring \cp-odd observable. Measuring a non-zero value for such an observable would directly imply the presence of \cp violation. While proposals for \cp-odd observables targeting the top-Yukawa coupling exists, their measurement is experimentally challenging. \cp violation in the Higgs--top-quark interaction, however, also induces deviations from the SM in \cp-even observables. While a deviation from the SM in a \cp-even observable is not guaranteed to be caused by \cp violation, a systematic investigation of indirect constraints is still a powerful method to narrow down the available parameter space for \cp violation.

In the present work, we perform a fit to derive bounds on a \cp-violating top-Yukawa coupling following the indirect approach. We take into account all available inclusive and differential Higgs boson measurements. Based upon the results of this fit, we point out that a measurement of single top quark associated Higgs production --- without relying on an assumption about the Higgs \cp character --- would significantly enhance the sensitivity to a \cp-violating top-Yukawa coupling. We then propose a strategy for performing such a measurements at the HL-LHC focusing on the Higgs boson decay to two photons.


\section{Top-associated Higgs production as a probe of the top-Yukawa interaction}
\label{sec:topH_prod}

We parameterize BSM effects in the Higgs-boson interaction with top quarks in the form
\begin{align}\label{eq:topYuk_lagrangian}
\mathcal{L}_\text{yuk} = - \frac{y_t^\SM}{\sqrt{2}} \bar t \left(\ct + i \gamma_5 \cttilde\right) t H,
\end{align}
where $y_t^\SM$ is the SM top-Yukawa coupling, \ct rescales the \cp-even top-Yukawa coupling ($\ct = 1$ in the SM), and \cttilde constitutes a \cp-odd top-Yukawa coupling ($\cttilde = 0$ in the SM). Additionally, we introduce the parameter \cv rescaling the Higgs couplings to massive vector bosons. To parameterize the effect of additional BSM particles affecting the $H\to \gamma\gamma$ and $gg\to H$ processes, we, moreover, float the Higgs--gluon--gluon and the Higgs--photon--photon couplings as free parameters (denoted as \kg and \kgamma, respectively).

The most relevant processes to constrain the modified top-Yukawa coupling of Eq.~(\ref{eq:topYuk_lagrangian}) are Higgs production via gluon fusion, the Higgs decay to two photons, $Z$-boson associated Higgs production, and top-associated Higgs production. Top-associated Higgs production is of special interest since the top-Yukawa coupling appears at the tree level allowing for a comparably model-independent probe of the Higgs--top-quark interaction.  Three different sub-processes contribute to top-associated Higgs production: $t\bar tH$, $tWH$, and $tH$ production. While $tWH$ and $tH$ production are negligible in the SM, their cross-section can be significantly enhanced in the presence of a \cp-violating contribution to the top-Yukawa coupling. We do not consider constraints arising from processes involving a virtual Higgs boson (e.g.\ $t\bar t$ or $t\bar t t \bar t$ production).

We perform the global fit of this model to all relevant Higgs boson measurements --- including the latest Higgs rate measurements from ATLAS and CMS as well as the $p_T$-binned simplified template cross-section measurements for the process $pp\to Z H, H\to b \bar b$ --- using \texttt{HiggsSignal}\cite{Bechtle:2013xfa,Bechtle:2014ewa,Bechtle:2020uwn}.

\begin{figure}\centering
\begin{minipage}{0.45\textwidth}
\includegraphics[width=\textwidth]{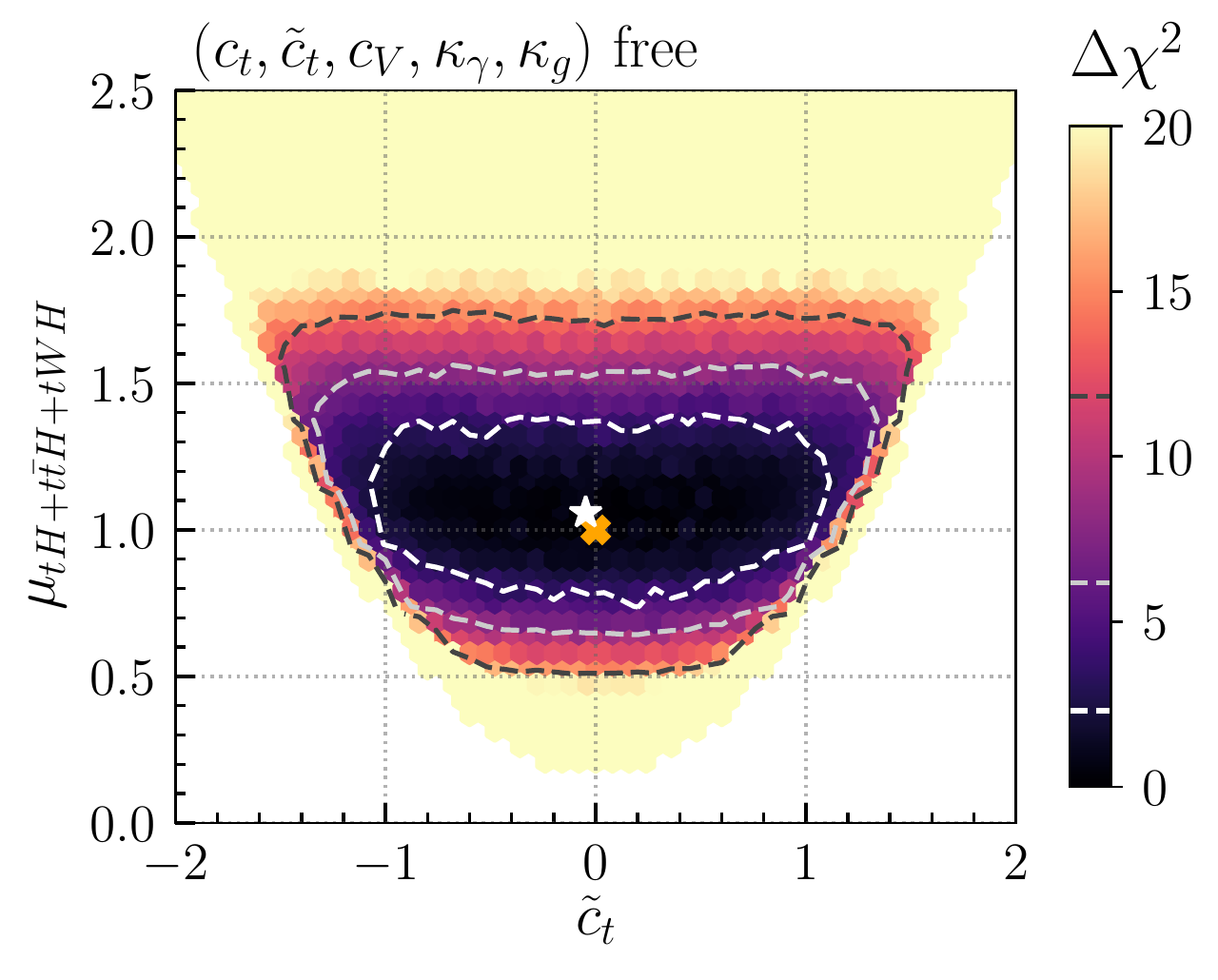}
\end{minipage}
\begin{minipage}{0.45\textwidth}
\includegraphics[width=\textwidth]{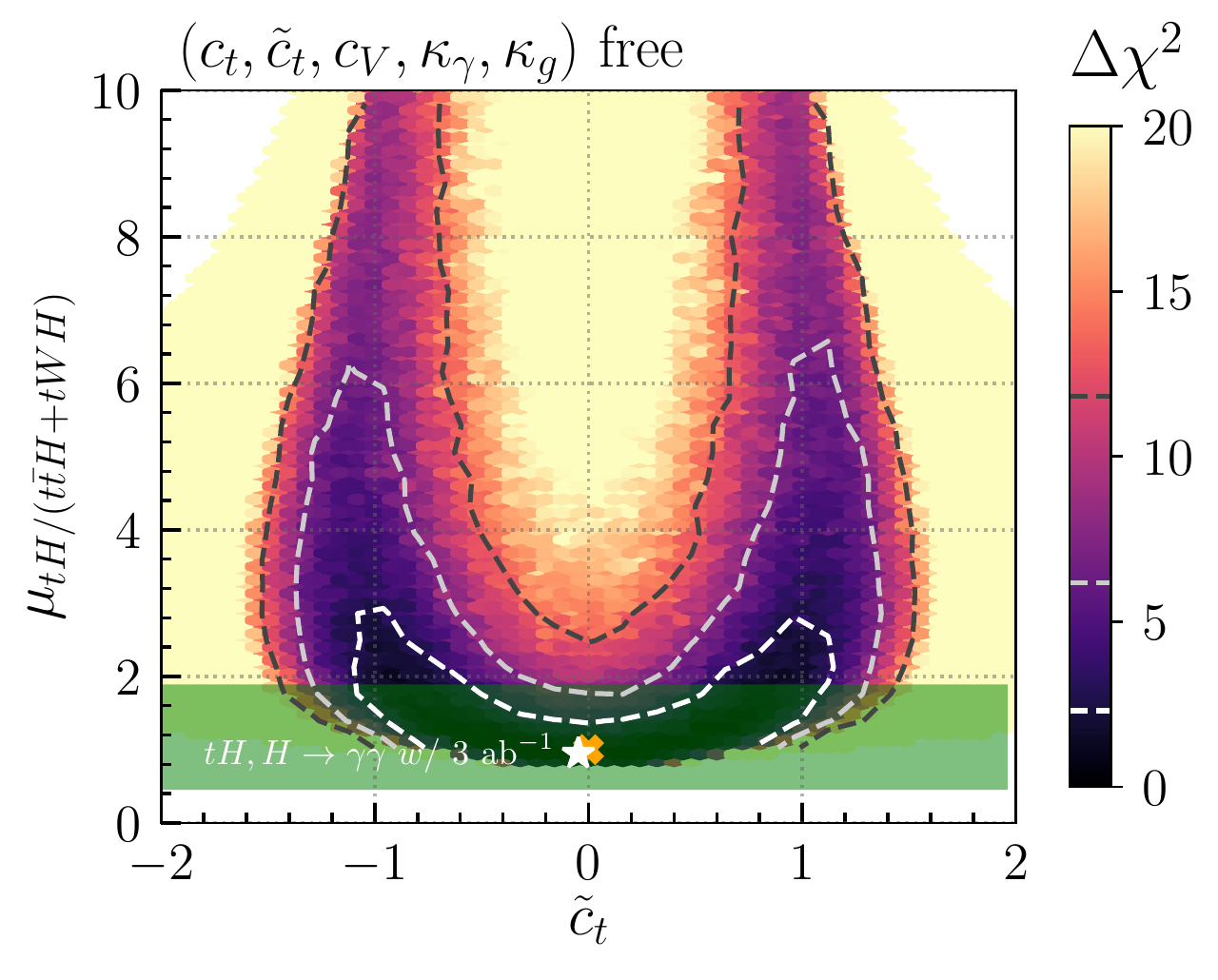}
\end{minipage}
\caption{\textit{Left:} Signal strength for combined top-associated Higgs production for the five-dimensional global fit in dependence of \cttilde. The color encodes the profiled $\Delta\chi^2$ distribution of the fit. The white, light-gray, and dark-gray contour lines represent the $1\,\sigma$, $2\,\sigma$, $3\,\sigma$ confidence regions, respectively. \textit{Right:} Same as left plot but the signal strength for $tH$ Higgs production divided by the signal strength for $t\bar t H$ and $tWH$ production is shown. The green band indicates the impact of the projected $tH$ measurement with $3\,\text{ab}^{-1}$ at the HL-LHC.}
\label{fig:topH_signal_strengths}
\end{figure}

In Fig.~\ref{fig:topH_signal_strengths}, we show two exemplary results of the described fit focusing on top-associated Higgs production. Since the different top-associated Higgs production channels are hard to distinguish experimentally, often a combination of these is measured. In the left plot of Fig.~\ref{fig:topH_signal_strengths}, the cross section for the sum of all three channels (normalized to the SM prediction) is shown as a function of \cttilde as predicted by the fit described above. The dependence of $\mu_{tH+t\bar tH+tWH}$ on \cttilde is approximately flat indicating that even a future more precise measurement of $\mu_{tH+t\bar tH+tWH}$ will not allow to tighten the constraints on \cttilde. If it would instead be possible to disentangle $tH$ production from $t\bar tH$ and $tWH$ production, the bounds on \cttilde could be improved. This is evident in the right plot of Fig.~\ref{fig:topH_signal_strengths} displaying the SM-normalized $tH$ over $t\bar t H + tWH$ cross section ratio as a function of \cttilde.


\section{Measuring single-top-associated Higgs production}
\label{sec:tH_measurement}

As discussed in Sec.~\ref{sec:topH_prod}, disentangling $tH$ and $t\bar tH + tWH$ production is an important step to tighten constraints on a \cp-violating Higgs--top-quark interaction. One possible strategy is to exploit the different lepton multiplicities of $tH$ and $t\bar tH + tWH$ production~\cite{Bahl:2020wee} by defining a one-lepton and a two-lepton category. Since only $t\bar t H + tWH$ production can contribute to the two-lepton category, the cross section in the two-lepton category can be used as a $t\bar tH + tWH$ control measurement for the one-lepton category to which all three sub-channels contribute. To enhance the sensitivity of this analysis, the one-lepton category is optimized for a high $tH$ event fraction. At this point it is important to make sure that such a $tH$ rate measurement is applicable for constraining the \cp nature of the Higgs boson. I.e., it should be ensured that the measurement is independent of the Higgs \cp nature (because otherwise the measurement could not be used to constrain the Higgs \cp nature).

\begin{figure}\centering
\begin{minipage}{0.42\textwidth}
\includegraphics[width=\textwidth]{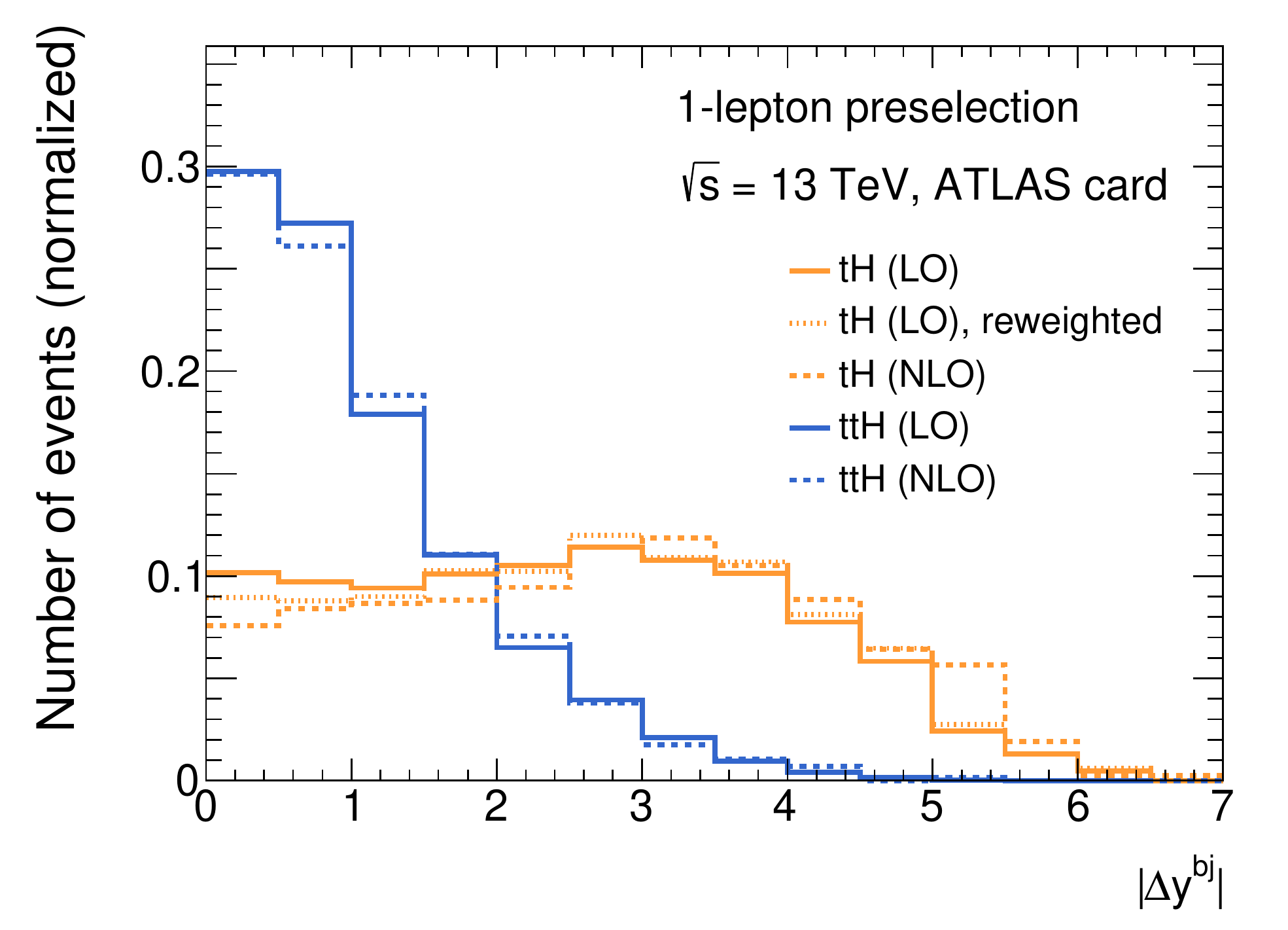}
\end{minipage}
\begin{minipage}{0.42\textwidth}
\includegraphics[width=\textwidth]{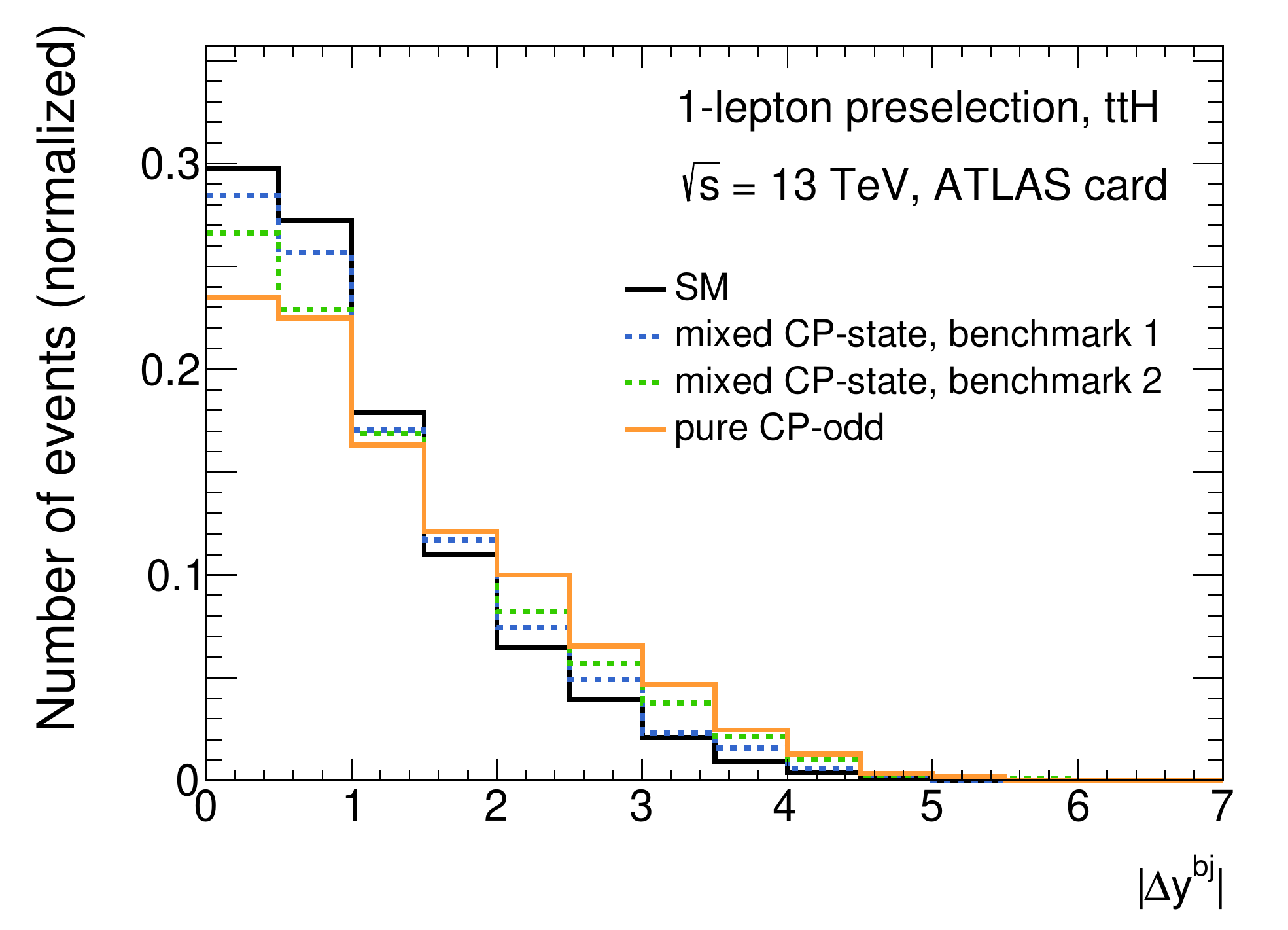}
\end{minipage}
\begin{minipage}{0.42\textwidth}
\includegraphics[width=\textwidth]{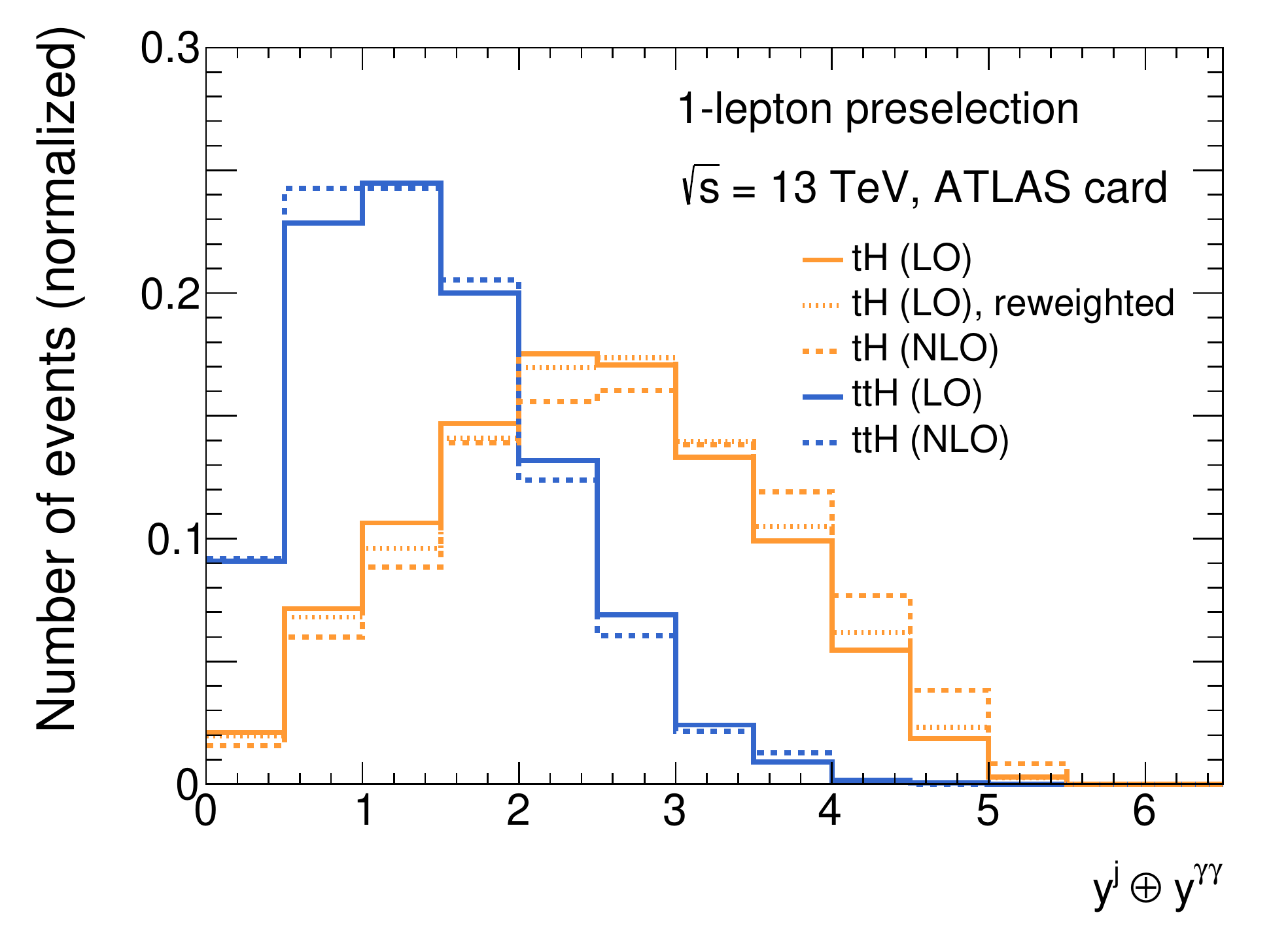}
\end{minipage}
\begin{minipage}{0.42\textwidth}
\includegraphics[width=\textwidth]{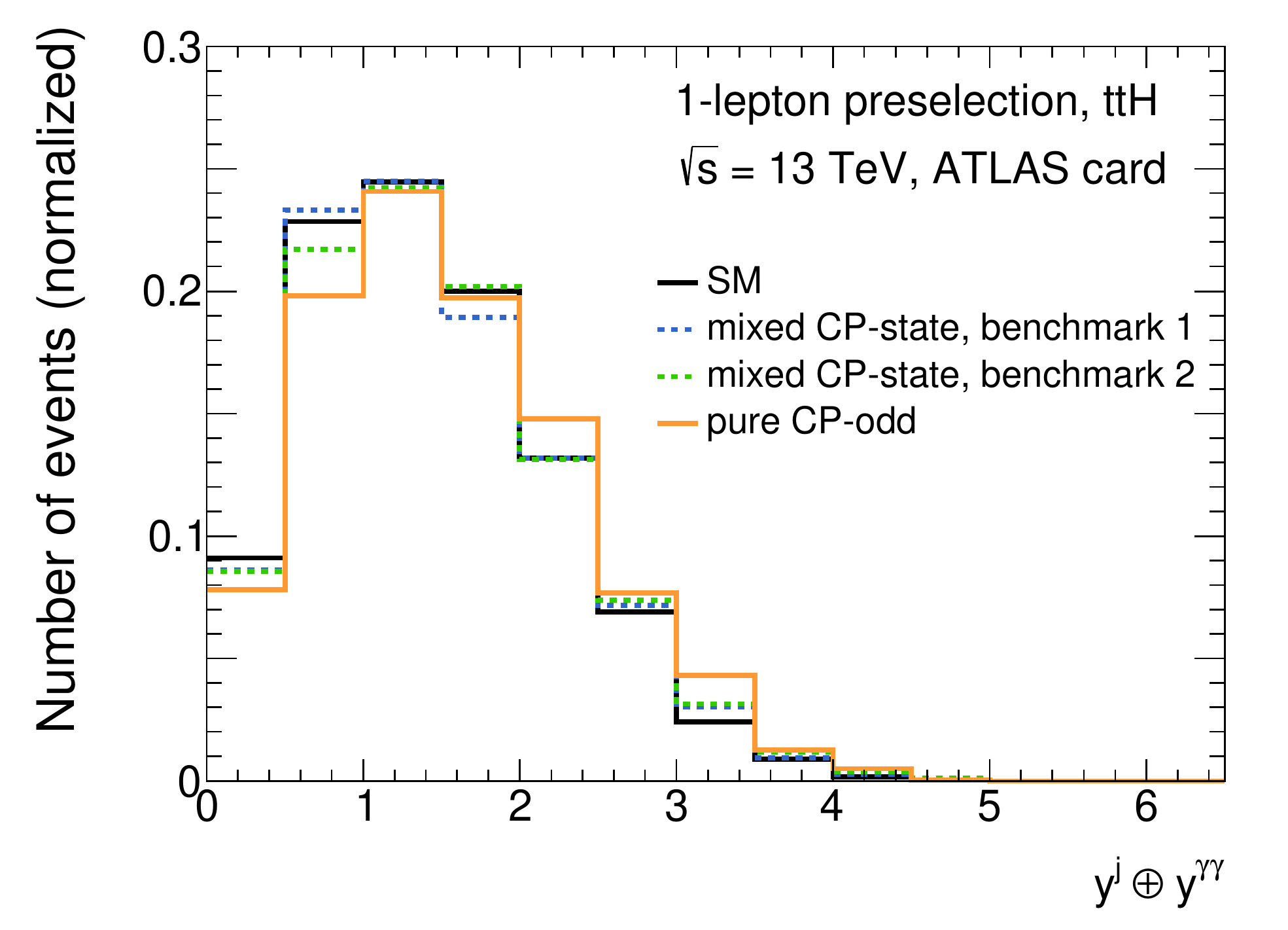}
\end{minipage}
\caption{\textit{Upper left:} $|\Delta y^{bj}|$ shape comparison of $t\bar t H$ (blue) and $tH$ (orange) production. \textit{Lower left:} Same as upper left panel, but $y^j \oplus y^{\gamma\gamma}$ is shown. \textit{Upper right:} $|\Delta y^{bj}|$ shape of $t\bar t H$ production comparing different \cp hypotheses. \textit{Lower right:} Same as upper right panel, but $y^j \oplus y^{\gamma\gamma}$ is shown.}
\label{fig:topH_event_shapes}
\end{figure}

The difficulty of this is indicated in Fig.~\ref{fig:topH_event_shapes} for the $H\to\gamma\gamma$ decay channel.  While e.g.\ the rapidity difference between the leading $b$ jet and the leading non-$b$ jet $|\Delta y^{bj}|$ is a good discriminator between $tH$ and $t\bar tH+tWH$ production (see upper left panel), cuts based on this observable introduce a large dependence of the measurement on the \cp character of the top-Yukawa coupling of $\sim 40\%$ (see upper right panel). Using instead the geometric mean of the leading jet and the Higgs boson rapidities $y^j \oplus y^{\gamma\gamma}$, $tH$ and $t\bar tH+tWH$ production can be disentangled equally well (see lower left panel) with a negligible ($\lesssim 2\%$) dependence on the Higgs \cp character (see lower right panel).

This strategy has been used to obtain a projection for a $tH$ measurement at the HL-LHC \cite{Bahl:2020wee}: the $tH$ signal strength is projected to be below $2.21$ at 95\% CL assuming SM-like data and using $3\,\text{ab}^{-1}$ (a limit five times stronger than the current strongest limit\cite{ATLAS:2020ior}). The constraint on \cttilde which would be imposed by such a measurement (for SM-like data) is indicated by the green band in the right plot of Fig.~\ref{fig:topH_signal_strengths}.


\section{Conclusion}

In the present article, we investigated how a \cp-violating Higgs--top-quark interaction can be constrained indirectly at the LHC. Using all relevant inclusive and differential Higgs-boson measurements, we performed a global fit showing that a more precise determination of top-associated Higgs production combining the $tH$, $t\bar tH$, and $tWH$ channels would not result in improved bounds on a \cp-odd top-Yukawa coupling. In order to improve the sensitivity, it will be crucial to disentangle $tH$ from $t\bar tH+tWH$ production without relying on assumptions on the Higgs \cp character.

We then discussed a potential strategy for reaching this goal by defining signal regions with different lepton multiplicities. We showed that it is especially important for the optimization of the analysis cuts to always keep the dependence of the kinematic acceptances on the Higgs \cp character as low as possible. Using $3\,\text{ab}^{-1}$, we projected an upper limit on the $tH$ signal strength five times strong than the current strongest limit.


\section*{Acknowledgments}

I would like to thank Philip Bechtle, Sven Heinemeyer, Judith Katzy, Tobias Klingl, Krisztian Peters, Matthias Saimpert, Tim Stefaniak, and Georg Weiglein for collaboration on the work presented here.


\paragraph{Funding information}

H.B. acknowledges support by the Deutsche Forschungsgemeinschaft (DFG, German Research Foundation) under Germany’s Excellence Strategy — EXC 2121 “Quantum Universe” — 390833306.

%
%

\bibliography{bibliography.bib}

\nolinenumbers

\end{document}